\documentclass[10pt]{iopart}

\usepackage{iopams} 
\expandafter\let\csname equation*\endcsname\relax
\expandafter\let\csname endequation*\endcsname\relax
\usepackage{amsmath}  
\usepackage{graphicx}
\usepackage{caption}
\usepackage{subfig}
\usepackage{color}
\usepackage[margin=1in]{geometry}
\usepackage{multirow}
\usepackage{tabularx}
\usepackage{cite}

\begin{document}

\title{Implications of the Klein tunneling times on high frequency graphene devices using Bohmian trajectories}

\author{Devashish Pandey$^{1}$, Matteo Villani$^{1}$, Enrique Colom\'{e}s$^{1}$, Zhen Zhan$^{2}$ and Xavier Oriols$^{1}$}

\address{$^{1}$ Departament d\rq{}Enginyeria Electr\`{o}nica, Universitat Aut\`{o}noma de Barcelona, Spain.\\
              $^{2}$ School of Physics and Technology, Wuhan University, Wuhan 430072, China\\}
\ead{xavier.oriols@uab.es}
\vspace{10pt}
\begin{indented}
\item[\today]
\end{indented}

\begin{abstract}
Because of its large Fermi velocity, leading to a great mobility, graphene is expected to play an important role in (small signal) radio frequency electronics. Among other, graphene devices based on Klein tunneling phenomena are already envisioned. The connection between the Klein tunneling times of electrons and cut-off frequencies of graphene devices is not obvious. We argue in this paper that the trajectory-based Bohmian approach gives a very natural framework to quantify Klein tunneling times in linear band graphene devices because of its ability to distinguish, not only between transmitted and reflected electrons, but also between reflected electrons that spend time in the barrier and those that do not. Without such distinction, typical expressions found in the literature to compute dwell times can give unphysical results when applied to predict cut-off frequencies. In particular, we study Klein tunneling times for electrons in a two-terminal graphene device constituted by a potential barrier between two metallic contacts. We show that for a zero incident angle (and positive or negative kinetic energy), the transmission coefficient is equal to one, and the dwell time is roughly equal to the barrier distance divided by the Fermi velocity. For electrons incident with a non-zero angle smaller than the critical angle, the transmission coefficient decreases and dwell time can still be easily predicted in the Bohmian framework. The main conclusion of this work is that, contrary to tunneling devices with parabolic bands, the high graphene mobility is roughly independent of the presence of Klein tunneling phenomena in the active device region. 
\end{abstract}


\ioptwocol

\section{Introduction}
\label{intro}

Because of its extraordinary properties, graphene has been studied as a new and promising material for electronics during the last fifteen years\cite{others31}. Although the lack of bandgap makes its use difficult  for digital applications, its high mobility is expected to provide very well suited devices for (small signal) radio frequency applications \cite{others23}. For electrons in graphene, modeled by the Dirac equation (with linear bands), an exotic tunneling phenomena, known as Klein tunneling \cite{klein, Allain_Fuchs}, is predicted resulting in a perfect transmission of electrons perpendicular to a potential barrier. This result is in contradiction with traditional semiconductors with parabolic bands where the transmission strongly depends on the height and width of the barrier. Several prototypes have already been studied in the literature for developing graphene field effect transistors for high-frequency applications based on Klein tunneling phenomena \cite{dev1,dev2,dev3,dev4}. The natural question arises: what is the mobility of electrons in graphene when undergoing Klein tunneling? This question is directly related to the time spent by the electrons in the region where they suffer Klein tunneling. For the sake of simplicity, hereafter, we assume the device active region to be equal to the potential barrier in graphene, meaning that transit and tunneling times are equivalent. The mobility determines the electron transit time that, in turn, determines the cut-off frequency which is an important figure of merit of high-frequency electron devices \cite{FoM}. Surprisingly, the connection between these Klein tunneling times and cut-off frequencies of graphene devices remains mainly unstudied in the literature. 

The accurate prediction of tunneling times has been a fascinating problem for the scientific community during the last century. In (non-relativistic) quantum mechanics, time enters as a parameter rather than an observable. Thus, there is no direct way to calculate tunneling times in the orthodox quantum mechanics, where measurements are directly linked to operators of the measured property \cite{tunreview}. The tunneling times can be indirectly determined by measuring other operators. In the literature, there exists at least three different orthodox protocols to compute the tunneling time \cite{others19}.  First, one studies the evolution of the wave packets through the barrier and gets the phase time which involves the phase sensitivity of the tunneling amplitude to the energy of the incident particle \cite{others20}. The second approach makes use of a physical clock to measure the time elapsed during the tunneling \cite{others22, others80,  others24, others25, others26}. Larmor precession, as one of physical clocks, was first introduced long time ago to measure the time associated with scattering events \cite{others80, others25}. Recently, tunneling times of 2D massless pseudo-spin Dirac particles have been analyzed, mainly within the second protocol \cite{others1,others2,others3, others14, others15, others16, others17,others18}. Finally, the third definition of tunnel time is based on the determination of a set of dynamic paths. We will refer to such type of tunneling time as the dwell time. However, a dynamic path is an ill-defined concept in orthodox quantum mechanics \cite{others21}.

In this paper, we will show that the Bohmian explanation of quantum phenomena provides a very appropriate formalism for discussing tunneling times that are later linked to cut-off frequencies. The Bohmian theory allows an accurate definition of dynamic paths (in terms of Bohmian trajectories) and the third alternative mentioned above for computing tunneling times becomes very natural. The most important advantage of the Bohmian computation of the dwell time for high-frequency electronics is its ability to distinguish, not only between transmitted and reflected electrons in the barrier \cite{leavens}, but also between those reflected particles that spend some time in the barrier and other reflected particles that do not spend time in the barrier. 

The structure of the paper is the following. In \sref{sec:2bis} we present how the transit (tunneling) time is related to the cut-off frequency of an electron device, specifically in graphene devices. In \sref{sec1} we define the dwell time from an orthodox perspective and from the Bohmian theory. In \sref{sec3}, we explain how the different dwell times can be computed. For that purpose, the Klein tunneling effect is presented and analyzed. Numerical results are shown and discussed in \sref{sec4}. Finally, we conclude in \sref{sec5}.

\section{Cut-off frequency and tunneling times}
\label{sec:2bis}

Along the paper we will consider a graphene two-dimensional (2D) sheet,  with  $x$ as the transport direction, from the left contact to the right contact, and $z$ as the direction perpendicular to the transport direction. The $y$ direction contains the thickness of the graphene sheet (plus top and bottom dielectric layers). We discuss in this section how the high mobility of graphene devices can be determined from the transit times. To simplify the discussion, we focus on a two terminal device. The length of the device active region is $L_x=b-a$ with $x=a$ the position of the left metallic contact and $x=b$ the right metallic contact. 

At very high frequencies, not only the particle current due to movements of particles is relevant, but also the displacement current given by the time-derivative of the electric field generated by electrons moving inside the device region becomes important. If we consider that the lateral surfaces of the metallic contacts ($L_y \times L_z$) are much larger than the length of the device, $L_x << L_y, L_z$, then, the Ramo-Shockley-Pellegrini theorem \cite{ramo,shockley,pellegrini}, allows us to write the total (particle plus displacement) current at each time $t$ as: 
\begin{equation}
\centering
I(t)=\dfrac{q}{L_x}\sum_{i=1}^{N_e} v_x^i(t) \Theta[x^i(t)-a]\Theta[b-x^i(t)]
\label{2terminal}
\end{equation}
where $N_e$ is the number of electrons inside the active region at time $t$, $q$ is the electron charge and $v_x^i(t)$ the $i-th$ electron instantaneous velocity in the transport direction $x$. Finally, $\Theta[x]$ is the Heaviside function. Notice that the trajectories in the metals are not included in \eref{2terminal}. This result assumes that the density of electrons in the metal and their mobility are so high that the electrical field generated by one moving electron in the metal (outside the active region) is rapidly screened by the other (free) electrons in the metal. 

Now, we consider the transient of the current $I(t)$ defined in \eref{2terminal} after a sudden perturbation of the external bias in the contacts. We assume that before $t=0$, the device has fixed external voltages in the contacts ($V_L$ in the left and $V_R$ in the right) with a stationary current value \footnote{We neglect the fluctuations of the stationary value of the current because they are not relevant in our discussion.}. Then, we apply a new (small signal) external bias $V_R+\Delta V$ at time $t=0$ at the right contacts of the device. This new external voltage generates a new internal potential in the graphene sheet $V(x,z)$ which perturbs the current given by \eref{2terminal} because the dynamics of the electrons traversing the device need to be adapted to the new scenario. After some time, the current reaches a new stationary value when all electrons inside the system have already moved along the active region $a<x<b$, all the time, with the new internal potential profile associated to $V_R+\Delta V$. This transient time can be related to the dynamics of electrons. Let us consider one electron, labeled by $i$, that has entered inside the active region just before $t=0$. The electron gives a current $I(t)=q\; {v_x^i(t)}/{L_x}$ during the time $\tau_{i}=\int_{0}^{\infty}dt\; \Theta[x^i(t)-a]\Theta[b-x^i(t)]$ that it spent in the active region. After this time interval, we are sure that a new electron (with identical properties except the entering time) entering inside the region $a<x<b$ at time $t>\tau_i$ will be only influenced by the new scenario created by the external bias $V_R+\Delta V$. Notice that the time $\tau_i$ we have to wait is not related to the fact that the electron is transmitted of reflected. The  only relevant point is that the electron spends some time in the active region $a<x<b$.

In a real device, there are more than one electron with some uncertainties in their properties. We only have access to the probability distribution of these uncertainties. Therefore, we compute an average value of the time spent by electrons over all these uncertainties to get the (average) transit time $\tau$. In this work, as already indicated, we will simplify our discussion by considering an active region built from a graphene potential barrier between two metallic contacts. Then, the transit time along the device and the tunneling time can be considered equivalent.  Finally, the previous relationship between transient of the current and dynamics of electrons can be formally established in the following expression between the ensemble transit (tunneling) time $\tau$ and the cut-off frequency of the device $f_T$ as\footnote{Typically, the clock frequency of a real CPU is usually 1/3 of the cut-off frequency. In any case, such factor is not relevant at all in the discussion presented here.}:  
\begin{equation}
\centering
f_T=\dfrac{1}{ \tau }
\label{ft}
\end{equation}
The language used above in terms of trajectories, which is natural for classical systems, can also be rigorously extended to quantum systems by using Bohmian trajectories \cite{oriols}. 

\section{Definition of dwell times in graphene}
\label{sec1}

The dynamics of electrons in graphene devices (as well as for other linear band structures materials) are given by the Dirac equation, and not by the usual Schr\"{o}dinger equation for parabolic bands. The wave function associated to the electron is no longer a scalar, but a bispinor \footnote{Throughout the text we will represent the bispinor wavefunction as $\psi$, while the scalar wavefunction will be represented with subscript ($\psi_i$).}:
\begin{eqnarray}
\centering
{\psi(\vec{r},t)}   \equiv 
\begin{pmatrix}
\psi_1\\ 
\psi_2 
\end{pmatrix}
 \equiv 
\begin{pmatrix}
\psi_1(x,z,t)\\ 
\psi_2(x,z,t)
\end{pmatrix}
\label{bispinor}
\end{eqnarray}
The two components are solution of the mentioned Dirac equation:
\begin{equation}
\centering
i\hbar\dfrac{\partial\psi(\vec{r},t)}{\partial t}=-i\hbar v_f(\vec{\sigma}\cdot\vec{\triangledown}+V(\vec r))\psi(\vec{r},t)
\label{Dirac}
\end{equation}
where $\vec r=\{x,z\}$ and $\vec \nabla=(\frac{\partial }{\partial x},\frac{\partial }{\partial z})$ and the Pauli matrices \footnote{In the literature, usually, our Pauli matrix $\sigma_z$ in \eref{sigma} is defined as the $\sigma_y$. However, since in this discussion the sheet of graphene is defined in the plane $XZ$, our notation is different.} are: 
\begin{eqnarray}
\centering
\label{sigma}
\vec \sigma=(\sigma_x, \sigma_z)=\left(
\begin{pmatrix}
0 & 1 \\
1 & 0
\end{pmatrix} ,
\begin{pmatrix}
0 & -i \\
i & 0
\end{pmatrix}\right)
\end{eqnarray}
We remind that $v_f=10^6  m/s$ is the graphene Fermi velocity and $V(\vec r) \equiv V(x,z)$ the electrostatic potential. 

As discussed in the introduction, from the different orthodox definitions of the tunneling times, we will use in this paper the third definition related to dynamic paths\cite{dypath}. It is the most accepted one, usually referred to as the dwell time, and it is not contaminated by the measurement procedure. By writing the modulus of the bispinor as $| \psi(x,z,t)|^2=|\psi_1(x,z,t)|^2+|\psi_2(x,z,t)|^2$, the typical expression for the orthodox dwell time to quantify how much time a particle spends in a 2D spatial region limited by the boundaries $a<x<b$ and $-\infty<z<\infty$ is traditionally given by, 
\begin{equation}
\centering
\label{tun_orth}
\tau_{D}=\int_{0}^{\infty}dt\int_{a}^{b}dx\int_{-\infty}^{\infty}dz| \psi(x,z,t)|^2
\end{equation}
At this point, let us briefly mention what is the tunneling time problem. A classical measurement of the dwell time can be simply defined from the measurement of the time when the particle reaches the position $x=a$, plus a final measurement of the time when the particle reaches $x=b$. The time spent between the initial and final detection of the particle position will quantify the dwell time. However, since quantum mechanics is a contextual theory, the first (strong) measurement of the position will transform the initial wave function into an eigenstate of the position measurement. Then, the posterior evolution of such delta function can be quite different from the unmeasured function used in \eref{tun_orth}. The tunneling time problem is related with the difficulties of computing the dwell time without paying the price of dealing with a perturbed wave function because of the measurement. In the orthodox theory, such attempt is quite difficult because only measured properties can be obtained from the theory. 

However, there are other quantum theories which can tackle such problem in a different way. By construction, the Bohmian theory\cite{bohm52} has the ability of providing measured and unmeasured properties (for example, particle positions) for a quantum system. If we know how to relate measured and unmeasured properties in one experimental set-up (for example, a high-frequency measurement set-up defined in \cite{PRL damiano}) then the computation of the  unmeasured properties of the Bohmian trajectories can be very useful. In this work, we will use unmeasured Bohmian trajectories to discuss Klein tunneling times.  As discussed in \cite{PRL damiano}, the measured Bohmian trajectories will only provide a noisier description of the total current (associated with a weak measurement process).  

In the Bohmian theory for the Dirac equation\cite{Holland93}, each electron has a well-defined position at any time that is guided by the same orthodox bispinor given by \eref{bispinor}. Each experiment labeled by the super index $i=1,..,N$ uses the same bispinor, but different trajectories $x^i(t)$ and $z^i(t)$.  Such Bohmian trajectories are computed by time-integrating the velocity given by the bispinor: 
\begin{equation}
\centering
\vec{v}(\vec{r},t)=\dfrac{ v_f \;  \psi(\vec{r},t)^{\dagger}\vec{\sigma} \psi(\vec{r},t)}{|\psi(\vec{r},t)|^2}
\label{bvel}
\end{equation}
where $\vec{\sigma}$  is defined in \eref{sigma} and $v_f = 10^6 m/s$  is the graphene Fermi velocity. See \ref{GrTr} for the explicit computation of \eref{bvel}. The initial position of such trajectories at time $t$ are empirically inaccessible and given by the probability distribution \cite{oriols}:   
\begin{equation}
\centering
\label{qe}
|\psi(x,z,t)|^2=\lim_{N\rightarrow\infty} \frac{1}{N}\sum _{i=1}^{N}\delta [x-x^i(t)]\delta [z-z^i(t)]
\end{equation}
where $N$ is the number of experiments that we assume infinite (or large enough to correctly get the ensemble values). In fact, due to equivariant property of the bispinor and the associated trajectories, if Eq.\eref{qe} is satisfied at the initial time, then it is true at any other time.  For a review on Bohmian mechanics, you can see \cite{bohm52, Holland93, oriols, PRL damiano, bell66}. 

Rewriting the orthodox expression of the dwell time in Eq.\eref{tun_orth} within the Bohmian language provide us more insights into the dwell time and its unmeasured definition \cite{oriols}. Using \eref{qe} in \eref{tun_orth} we get:
\begin{equation}
\centering
\label{tun_boh1}
\tau_{D}=\lim_{N\rightarrow\infty}\frac{1}{N}\sum _{i=1}^{N}\left(\int_{0}^{\infty}dt\Theta[x^i(t)-a]\Theta[b-x^i(t)]\right)
\end{equation}
Now, we can rewrite Eq.\eref{tun_boh1} as an average over Bohmian dwell times $\tau^i$ associated to the different trajectories:
\begin{equation}
\centering
\tau_{D}=\lim_{N\rightarrow\infty}\frac{1}{N}\sum _{i=1}^{N}\tau^i
\label{tun_boh2}
\end{equation}
 where $\tau^i$ is defined as:
\begin{equation}
\centering
\tau^i=\int_{0}^{\infty}dt\Theta[x^i(t)-a]\Theta[b-x^i(t)]
\label{tun_bohnew}
\end{equation}
This is the dwell time associated to the $i$-th Bohmian trajectory inside the region $a<x<b$.  Notice that the spatial integral in the $z$ direction from $-\infty$ to $\infty$ in \eref{tun_orth} implies that we do not care about which is the $z$ position of the particle.

Up to now, our Bohmian discussion is just another way to exactly compute the orthodox dwell time in  \eref{tun_orth}. We can now further develop the Bohmian expression to realize about its ability to discuss the high-frequency performance of electron devices discussed in \sref{sec:2bis}. We divide the trajectories appearing in $\tau_D$ in Eq.\eref{tun_boh2} into the three types of trajectories:
\begin{itemize}
\item  ($T$-trajectories) Those Bohmian trajectories that enter into the barrier region through $x=a$ and leave through $x=b$ being finally transmitted. We define $N_T$ as the number of such trajectories. By construction, their $\tau^i$ is different from zero. 

\item ($R$-trajectories) Those Bohmian trajectories that enter into the barrier region through $x=a$ and leave through the same point $x=a$ because they are finally reflected. We define $N_R$ as the number of such trajectories. Again, their $\tau^i$ is different from zero. 

\item ($R^*$-trajectories) Those Bohmian trajectories that do not enter into the barrier region at any time. We define $N_{R^*}$ as the number of such trajectories. These trajectories are reflected trajectories, but different from the $R$-trajectories. Here, by construction, we have $\tau^i=0$.  
\end{itemize}
By construction, with the new definitions, we have $N=N_T+N_R+N_{R^*}$. Then, the dwell time $\tau_D$ in Eq.\eref{tun_boh2} can be rewritten as:
\begin{equation}
\centering
\label{tun_boh3}
\tau_{D}=\lim_{N\rightarrow\infty}\frac{1}{N}\left(\sum _{l=1}^{N_T}\tau^l+\sum _{m=1}^{N_R}\tau^m\right)
\end{equation}
From the above equation we can define the transmission time, $\tau_T$ and the reflection time, $\tau_R$ as follows:

\begin{equation}
\centering
\label{tun_bohm5}
\tau_{T}=\frac{1}{N_T}\sum _{l=1}^{N_T}\tau^l \mbox{~  and  ~} \tau_{R}=\frac{1}{N_R}\sum _{m=1}^{N_R}\tau^m
\end{equation}
So the overall expression of the dwell time can be written as follows,
\begin{equation}
\centering
\label{tun_bohm6}
\tau_{D}=P_T\;\tau_T+P_R\; \tau_R
\end{equation}
where we have defined the probabilities:
\begin{equation}
\centering
\label{Tb}
T\equiv P_T =\lim_{N \to \infty} \frac{N_T}{N}
\end{equation}
The computation of the transmission coefficient $T$ do not require the distinction between $N_R$ and $N_{R}^*$ since only the transmitted trajectories $N_T$ are relevant here. Identically, \begin{equation}
\centering
\label{Rb}
P_R=\lim_{N \to \infty} \frac{N_R}{N}
\end{equation}
Notice that the reflected probability $P_R$  is different from the reflection coefficient $R$, $P_R\neq R$, because the reflection coefficient requires including $N_R$ and $N_{R}^*$ in the numerator of  \eref{Rb}.

We further discuss the role of the ${R^*}$-trajectories. Because of these trajectories the previous probability definitions give $P_T+P_R\le1$. We require to add the additional probability $P_{R^*}=N_{R^*}/N$ to satisfy $P_T+P_R+P_{R^*}=1$. However, if we remember that the ${R^*}$-trajectories have a transit time equal to zero, $\tau_i=0$, then, the transit (tunneling) time of expression \eref{tun_orth} can be extremely misleading. If we get a scenario where $N_{R^*} \approx N$ then we get the unphysical result $\tau_D \approx 0$ in \eref{tun_boh3}, that implies a cut-off frequency going to infinite from \eref{ft}. This result is unphysical. The mistake appears because we have to eliminate the trajectories $N_{R^*}$ from the computations of the dwell times when such times want to be related to predict the high-frequency behavior of electron devices as discussed in \sref{sec:2bis}. The fundamental problem is that the identification of the particles $N_T$, $N_R$ and $N_{R^*}$ is not possible within the orthodox theory. This is just a different way of realizing about the controversial tunneling time in orthodox quantum mechanics. On the contrary, the Bohmian theory provides a transparent procedure to eliminate $N_{R^*}$ from the computations. Thus, the Bohmian dwell time (for deducing properly high-frequency performances) needs to be defined as:
\begin{equation}
\centering
\label{tun_boh7}
\tau_{D_B}=\lim_{N_{B}\rightarrow\infty}\frac{1}{N_{B}}\left(\sum _{l=1}^{N_T}\tau^l+\sum _{m=1}^{N_R}\tau^m\right)
\end{equation}
where $N_{B}=N_T+N_R$ are the number of trajectories entering into the barrier. Notice that now the scenario $N_{R^*} \approx N$ does not imply the unphysical result $\tau_{D_B} \approx 0$ in \eref{tun_boh7} because the particles $N_{R^*}$ have no role. We will elaborate this point in the subsequent sections where numerical results are shown.

\section{Klein tunneling and the simulation set-up}
\label{sec3}

\begin{figure*}
\centering
\includegraphics[width=0.90\textwidth]{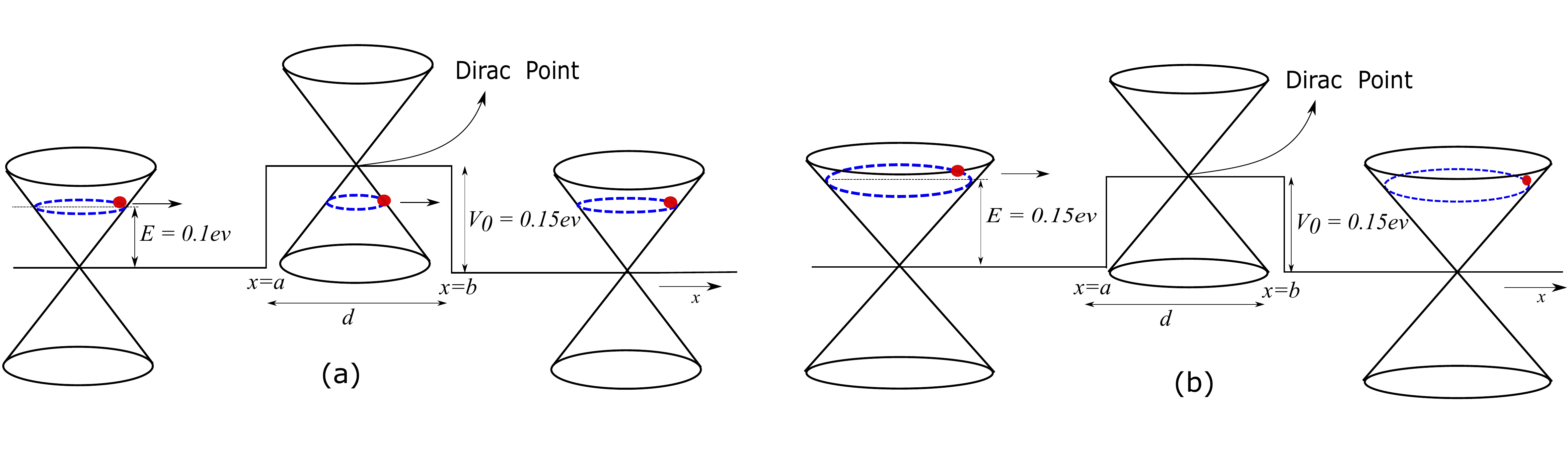}
\caption{(a) Klein tunneling barrier region where the electron, which impinges perpendicularly to the barrier, has an energy $E$ lower than the barrier height $V_0$. The cones represents the linear energy momentum dispersion at different positions. The electron has available states in the valence band of the barrier region which allows them to tunnel freely. The transmission coefficient in such cases is close to unity. (b) The same plot for an electron with energy similar to the barrier height $E=V_0$. In this case the electron has to occupy the Dirac point in the barrier region which has almost no available energy states. In these scenarios the transmission probability almost vanishes. This decrease can also be explained through a momentum conversation argument, as depicted in \fref{fig6}.}
\label{fig1}
\end{figure*}

Now we detail the quantum simulation of dwell times in graphene devices. All simulation results are done with the BITLLES simulator \cite{bits}. For the simulation, we consider a two terminal device whose band structure (energy of the Dirac point as a function of the $x$ position) is plotted in \fref{fig1}. We consider that the active region of the device is formed by the region with a barrier of $V_0=0.15$ eV that starts at the position $x=a=150$ nm and ends at $x=b=304$ nm. The simulation box in the $x$ direction is enlarged to be able to accommodate the central position of the initial wave packet at $x=0$ nm. The total simulation box in the $x$ direction is $1\mu$m, while, the one in the $z$ direction is $600$ nm. The spatial step for the computation of Dirac equation are $\Delta x=\Delta z= 1$ nm, while the time step is $\Delta t=10^{-5}$ ps. To simplify the discussion, the contact is assumed to have the same properties as the graphene channel, but with a very fast screening time so that the only electron relevant for the total current in \eref{2terminal} are the ones inside the box $a<x<b$. This scenario corresponds to the idealized two-terminal device described in \sref{sec:2bis}.

 As we mentioned, the wave nature of the electrons\footnote{The time-evolution of this wave packet can be considered as a Bohmian conditional wave function for the electron.  The conditional wave packet is a unique tool of Bohmian mechanics that allows to tackle the many-body and measurement problems in a computationally very efficient way \cite{OriolsPRL,EnriquePRB}.} is given by the Dirac equation using the bispinor in \eref{bispinor}. The initial electron wave function is a Gaussian bispinor wave packet:
\begin{equation}
\label{bispinorini}
\begin{pmatrix}
\psi_1(x,z,t) \\
\psi_2(x,z,t) 
\end{pmatrix}= \left(\begin{matrix}
1 \\ se^{i\theta_{\vec{k_c}}}
\end{matrix}\right)\psi_g(x,z,t)
\end{equation}
where $\psi_g(x,z,t)$ is a Gaussian function with central momentum $\vec{k_c}=(k_{x,c},k_{z,c})$. We consider  $s=1$ for wave functions with positive kinetic energies (conduction band) and $s=-1$ for negatives kinetic energies (valence band). We define $\theta_{\vec{k_c}}=arctan(k_{z,c}/k_{x,c})$ as the incident angle. The wavepacket spatial dispersions along the $x$ and $z$ directions are equal to $\sigma=40$ nm. Unless we indicate the contrary, the central energy of the electron will be $0.1$ eV above the Fermi point in the left contact.  

\begin{figure*}
\centering
\includegraphics[width=0.7\textwidth]{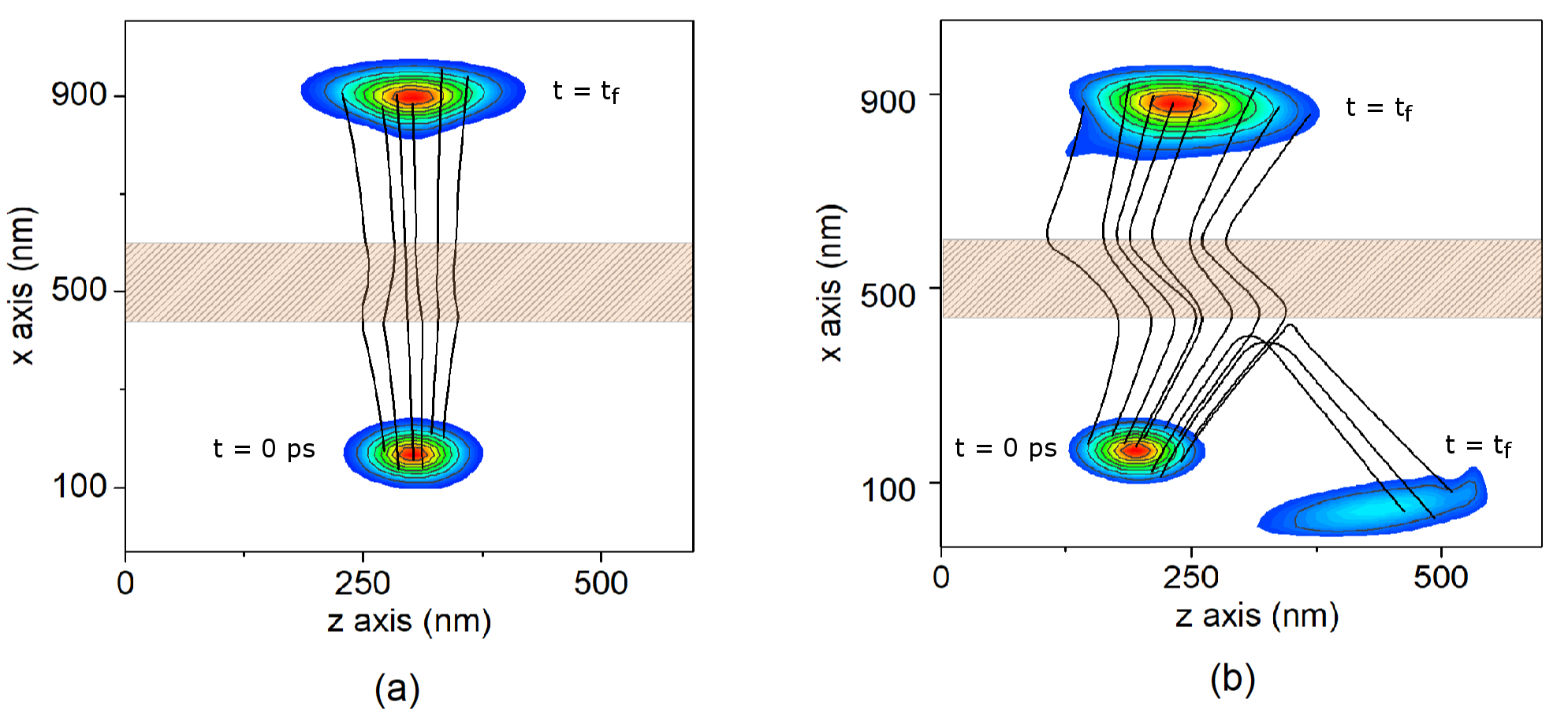}
\caption{(a) Conditional wavefunction of the electron that impinges perpendicularly ($\theta_{\vec{k_c}}=0$ degrees) to a barrier (in the shaded orange region) in the initial ($t=0$ ps) and final ($t_f=0.746$ ps) times. A set of the associated Bohmian trajectories are also plotted. As it can be seen, from both the wave packet and the set of trajectories, the electron exhibits Klein tunneling and all trajectories traverse the barrier. (b) The same plot for an electron that does not impinge perpendicular to the barrier ($\theta_{\vec{k_c}}=15$ degrees). Now, there is no complete Klein tunneling and part of the wave packet and some trajectories are reflected. The transmitted part of the wave packet and transmitted trajectories suffered refraction according to Snell's law-like expression \eref{snell}.}
\label{fig2}
\end{figure*}

We are interested in computing the time spent by an electron while traversing the potential barrier depicted in \fref{fig1} by  Klein tunneling \cite{nature}. For this purpose we simulated different scenarios where an electron (represented by its conditional wave function) impinges a barrier. In \fref{fig2}, we see two different examples of such conditional wave function and its associated trajectories\footnote{We observe in figure 2 (b) trajectories which are crossing in the $\{x,z\}$ space. We remind that different Bohmian trajectories cannot cross at the space $\{x,z,t\}$ at the same time. The trajectories plotted here satisfy this requirement.}. Let us notice that, in our opinion, the word tunneling is misleading here. As plotted in the cones of \fref{fig1}(a), electrons in the contact have kinetic energies available above and below the Dirac point. Identically, electrons inside the potential barrier have energies available above and below the new Dirac point. Therefore, strictly speaking, even if we consider an electron with an incident kinetic energy below the potential energy in the barrier, $E<V_0$, the electron will not find a region of forbidden energies as it happens in typical tunneling barriers built from materials with parabolic bands and with an energy gap. In this sense, the electron transport in the graphene linear band structure is quite unusual and unique. Rather than a tunneling phenomena is a interference phenomena. We will see these differences in next section and we will comment their implications in the conclusions.

We deduce here some features of the dynamics of the electrons traversing the barrier region. By construction, the total energy of the wave packet is conserved. Such energy can be divided between kinetic and potential energy. If we locate the zero of potential energy at the Dirac point in the left contact, then the electron has a positive kinetic energy of $E$. Once inside the barrier, the potential energy $V_0$ is higher than the total energy $E$, so that the kinetic energy in the barrier is negative $E-V_0$. These negative kinetic energies are unproblematic and perfectly well defined in the linear band structure of graphene.  If the initial kinetic energy of the incident electron is $E=V_0$, then, a quite exotic situation appears because there is almost no energy eigenstates available in the barrier region to accommodate the energy eigenestates that build the wave packet outside the barrier. See \fref{fig1}(b). The previous arguments are strictly valid for wave functions that contain just one energy eigenstate, for wave packets built from a set of energies around the central value, as in our case, the time evolution is more complex. 

\begin{figure}
\centering
\includegraphics[scale=0.6]{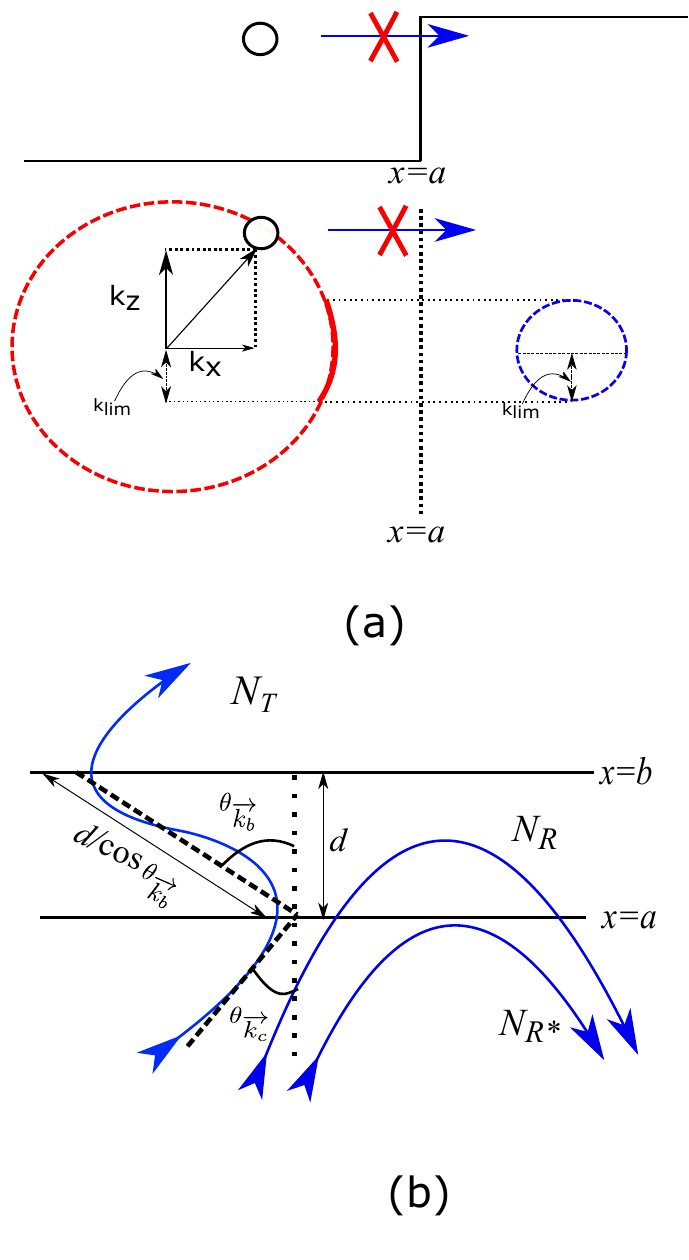}
\caption{(a) Scheme of an electron that cannot tunnel through the barrier region because of conservation of the total energy and the $z$ momentum forbids it. The electron is reflected before entering into the barrier, contributing to the $N_{R^*}$ particles, because there are no available $z$-momentum for the the corresponding kinetic energy in the barrier region. (b) Scheme depicting the three possible types of trajectories considered in this work: transmitted particles, $N_T$, particles entering into the barrier but eventually reflected, $N_R$ and  particles that are reflected before entering the barrier $N_{R^*}$. }
\label{fig6}
\end{figure}

When electrons are incident at a particular angle $\theta_{\vec{k_c}}=arctan(k_{z,c}/k_{x,c})$ with momentum $\vec{k_c}=(k_{x,c},k_{z,c})$, due to the translational invariance of the potential $V(x,z)$ in the $z$ direction, the momentum in that direction should be conserved. This means that, for example, the transport process depicted in \fref{fig6}(a) is not possible because the $k_{z,c}$ is not conserved. The $k_z$ in the left contact is much larger than the $k_{lim}$ value at the barrier region. The argument of conservation of $z$-momentum leads to scenarios where electrons change its direction in the interfaces contact-barrier and barrier-contact resulting in a Snell's law-like expression \cite{Allain_Fuchs}:
\begin{equation}
E\sin(\theta_{\vec{k_c}})=(E-V_0)\sin(\theta_{\vec{k_b}})
\label{snell}
\end{equation}
where $\theta_{\vec{k_c}}$ is the angle before the barrier (incident angle) and $\theta_{\vec{k_b}}$ the angle in the barrier (refracted angle). See \fref{fig6}(b) for a definition of the angles in the contact-barrier interface. In the barrier region, we have $E-V_0<0$ so that the angle of transmission of the trajectory is negative. From \eref{snell}, we conclude \cite{Allain_Fuchs} that the angle of incidence $\theta_{\vec{k_c}}$ in graphene has to satisfy $\theta_{\vec{k_c}} > \theta_C$ to have a completely reflected wavefunction, where
\begin{equation}
\theta_C=\sin^{-1}\left(\frac{E-V_0}{E}\right)
\label{C}
\end{equation}
is the critical angle. Again, the previous arguments are strictly valid for just one energy eigenstate. For wave packets built from a set of energies around the central value, as in our case, there can also  be a small transmission above the critical angle. 

The most surprising result for the dynamics of electrons in graphene, as already indicated, appears when considering an incident angle $\theta_{\vec{k_c}}=0$ meaning that the momentum in the $z$ direction is zero, $k_{z,c}=0$. Then, the conservation of the $z$ momentum does not provide any restriction on the dynamics of the electron and, in fact, the transmission coefficient is equal to one ($T=1$) for any positive or negative kinetic energy of the electron incident on the barrier with $\theta_{\vec{k_c}}=0$. This is known as the Klein tunneling paradox \cite{klein, nature} because it is surprising for typical tunneling (parabolic band) scenarios with forbidden energy regions. But, this is not the case in graphene, and the paradox just disappears. In our study, since the injected wave function is a wave packet, it will have some wave vectors with some dispersion in the injecting angle around $\theta_{\vec{k_c}}=0$, and therefore $T\leq1$.

\section{Numerical results and discussion}
\label{sec4}
We consider here a wave packet in \eref{bispinorini} with a kinetic energy given by $E=0.1$ eV located initially at the left side,  $x=0$ nm, far from the barrier region. We will consider different incident angles $\theta_{\vec{k_c}}$ that determine different propagation directions, meaning different $x$ and $z$ momenta $\{k_{x,c},k_{z,c}\}$. The time evolution of such bispinor, while traversing the barrier, is given by \eref{Dirac}.  From the knowledge of the bispinor at any time and position, we compute the velocity of the Bohmian trajectories from \eref{bvel}. By time integrating these velocities, we compute the Bohmian trajectories $\{x^i(t),z^i(t)\}$ where the super index $i$ specifies different experiments that imply different initial positions of the particles selected according to \eref{qe}. 

\begin{figure}
\centering
\includegraphics[scale=0.3]{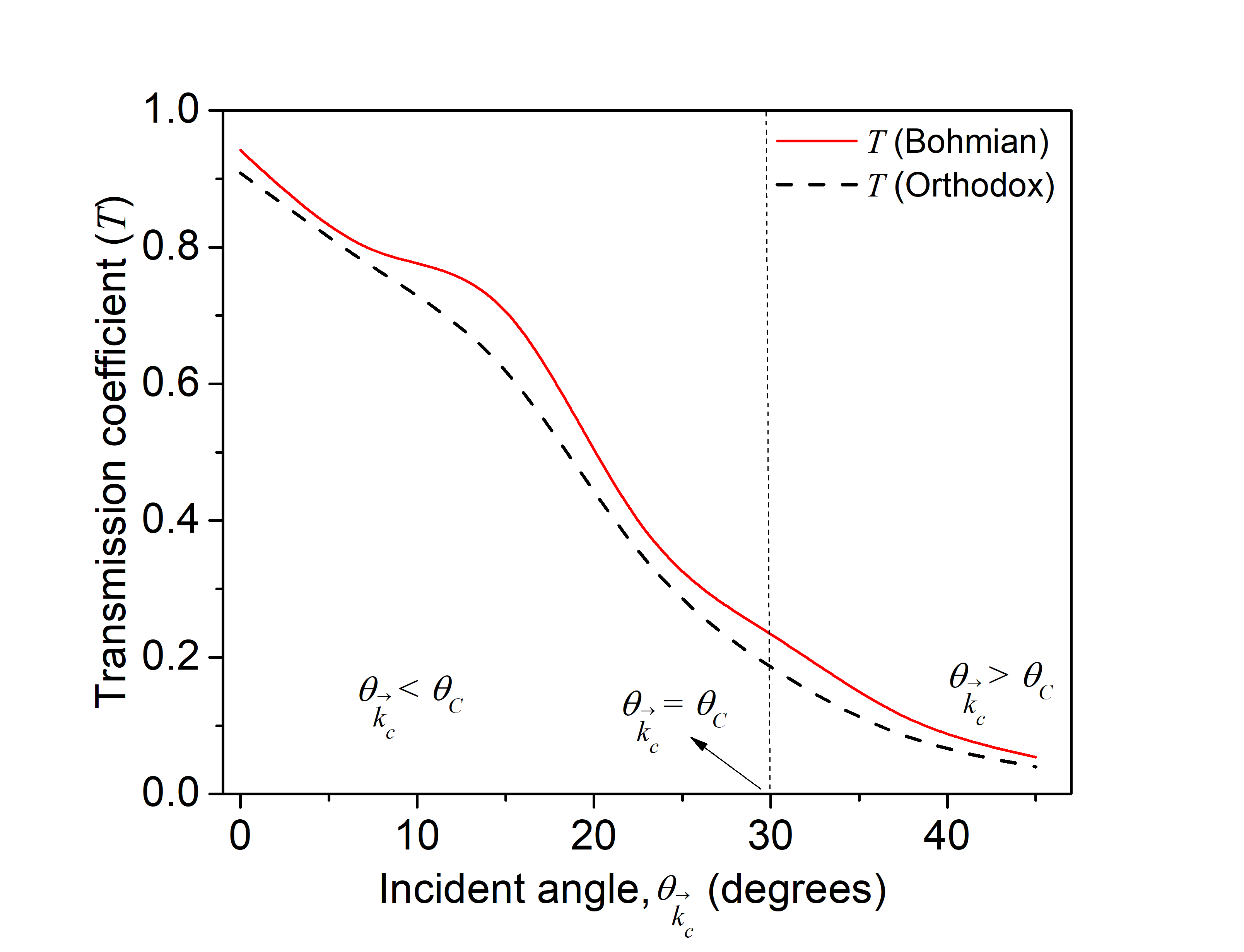}
\caption{Transmission coefficient as a function of the incident angle computed from the square modulus of the wave function in \eref{T} and the Bohmian trajectories in \eref{Tb}. Bohmian and orthodox computations show an excellent agreement.}
\label{fig10}
\end{figure}

First, we discuss the transmission coefficient that can be computed from the bispinor easily as:
\begin{equation}
\centering
\label{T}
T=\int_{b}^{\infty}dx\int_{-\infty}^{\infty}dz|\psi(x,z,t_{f})|^2
\end{equation}
where $t_{f}$ is a time large enough so that there is no probability presence in the barrier region. Identically, by putting \eref{qe} into \eref{T},  the transmission coefficient can be computed from the Bohmian trajectories as in \eref{Tb}. The plot in \fref{fig10} confirms that the results computed from the Bohmian trajectories in \eref{Tb}  (with $N$=500 experiments) reproduce accurately the orthodox results in \eref{T}. Following the discussion about the Klein tunneling in \sref{sec3}, for  $\theta_{\vec{k_c}}=0$ we get $T\approx 1$, while $T$ tends to zero as we increase the angle. We have a small transmission probability for  $\theta_{\vec{k_c}}=\theta_C$.  
\begin{figure}
\includegraphics[scale=0.3]{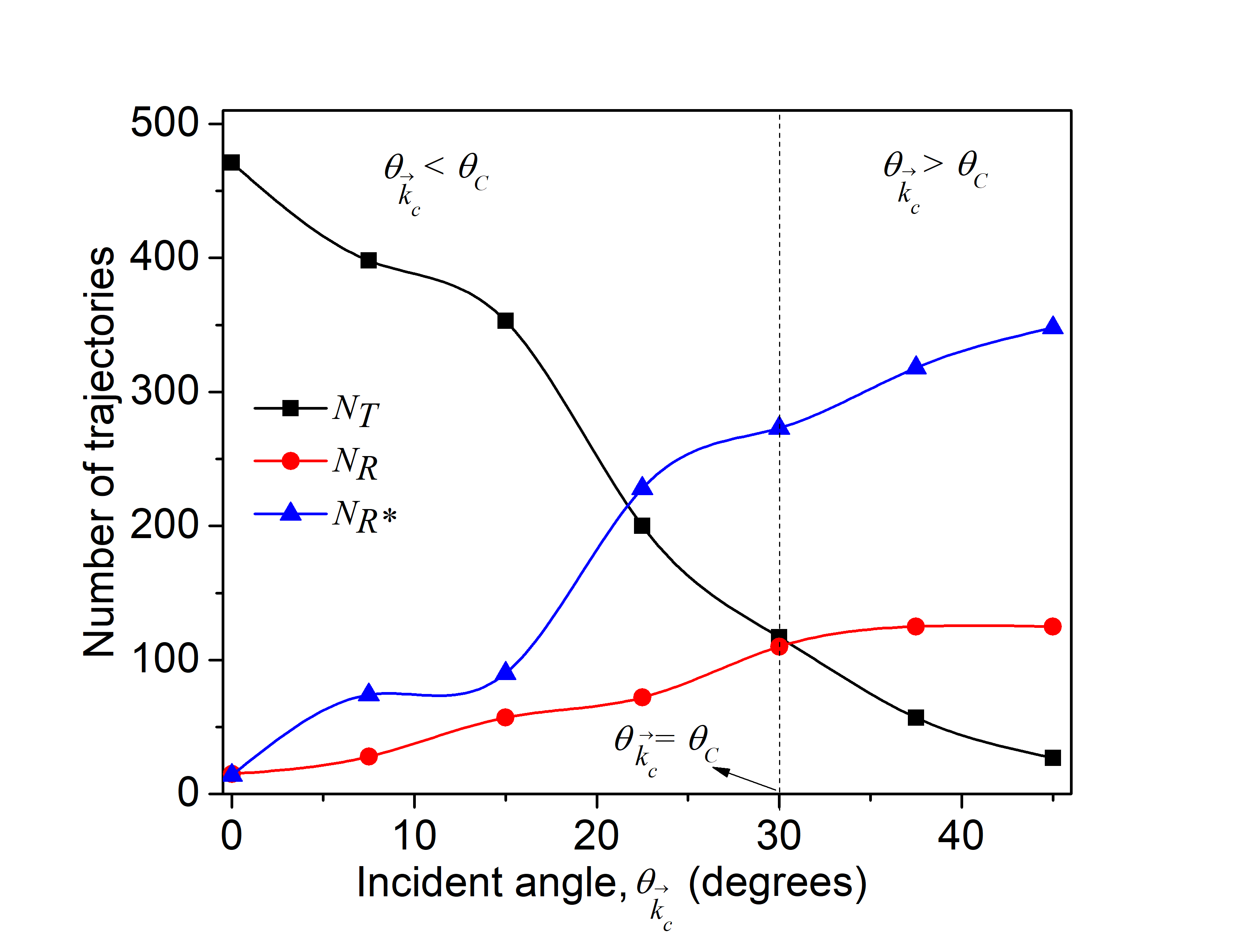}
\caption{Number of transmitted particles, $N_T$, particles entering into the barrier but eventually reflected, $N_R$ and  particles that are reflected before entering the barrier $N_{R^*}$ as a function of the incident angle.}
\label{fig3}
\end{figure}

As we discussed in \sref{sec1}, the correct computation of the dwell time requires the distinction among $N_T$, $N_R$ and $N_{R^*}$. With Bohmian mechanics it is possible to distinguish among the transmitted trajectories,  $N_T$ , reflected after entering in the barrier, $N_R$ and those that are reflected before entering the barrier, $N_{R^*}$.  The schematic representation of these trajectories is plotted in \fref{fig6}(b). In \Fref{fig3} we show how the number of these trajectories vary with the angle of incidence $\theta_{\vec{k_c}}$. The simulations show that for $\theta_{\vec{k_c}}=0$ almost all the particles are transmitted. Increasing  $\theta_{\vec{k_c}}$ leads to an increase in the reflected particles. By construction, the behavior of $N_T$ in \fref{fig3} just reproduces the transmission coefficient $T$ in \fref{fig10}. We divide these reflected Bohmian trajectories into two sets: $N_R$ and  $N_{R^*}$. The estimation of the current delay in \eref{2terminal} does only take into account  particles entering in the barrier, either $N_T$ or $N_{R}$. In the orthodox computation, just with the bispinor (without trajectories), $N_R$, $N_T$ and $N_{R^*}$ cannot be treated separately. This fact represents an important limitation of the orthodox theory in the proper description of tunneling times and, subsequently, high-frequency response of graphene devices.

\begin{figure}
\centering
\includegraphics[scale=0.3]{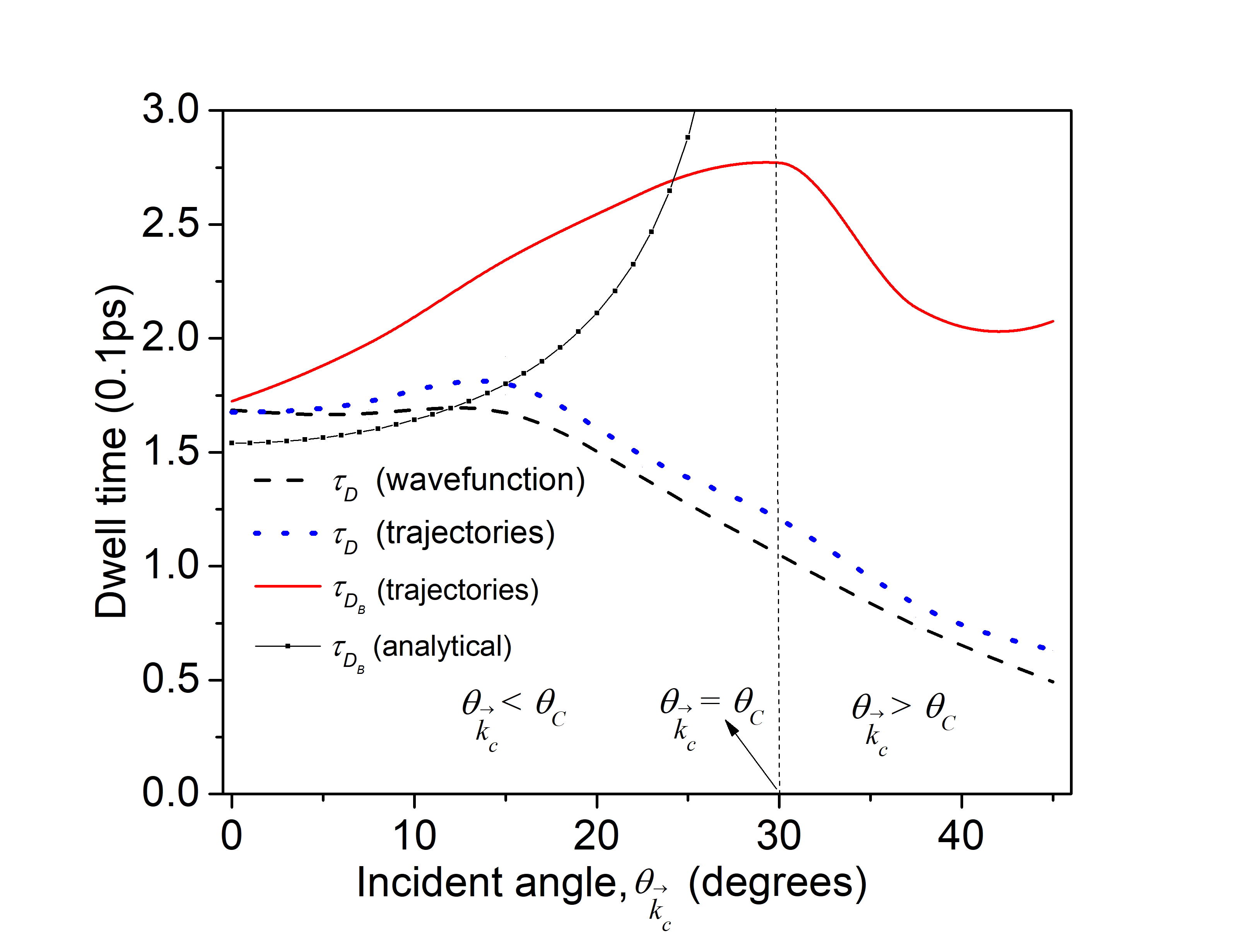}
\caption{Dwell time as a function of the incident angle computed from Eq.\eref{tun_orth} (black dashed line), Eq.\eref{tun_boh3} (blue dotted line), Eq.\eref{tun_boh7} (red solid line) and Eq.\eref{t2} (black solid line with square symbols).}
\label{fig5}
\end{figure}

In \fref{fig5}, we plot with dashed lines the orthodox dwell time $\tau_D$ given by \eref{tun_orth}. We find that it decreases monotonically with the increase of the incidence angle $\theta_{\vec{k_c}}$. These results are compatible with the decrease of the transmission coefficient $T$ in \fref{fig10} because particles have less and less probability to enter into the barrier region and, therefore, $\tau_D$ decreases. For an incident angle larger than the critical angle, $\theta_{\vec{k_c}}>\theta_C$, we expect $T \to 0$ and $\tau_D \to 0$. Then, using \eref{ft} for the computation of the cut-off frequency, we get an unphysical result of an infinite cut-off frequency $f_T \to \infty$. This unphysical result is also present in \eref{tun_boh1} computed with trajectories. The problem appears because of the large number of $N_{R^*}$ while $N_T, N_R \to 0$, at high incident angles (see \fref{fig10}). A physical computation of the dwell time can be obtained using the Bohmian trajectories in \eref{tun_boh7}, that ignores $N_{R^*}$, as seen in \fref{fig5} with solid line. This is one of the main results of this work. 

For $\theta_{\vec{k_c}}=0$, the situation is much more simple because $N_{R^*} \to 0$ and then the dwell time (either with the orthodox or Bohmian expression) is roughly equal to: 
\begin{equation}
\tau_{D}\approx \tau_{D_B} \approx \frac{d} {v_f}
\label{t1}
\end{equation}
Numerically, we get in \fref{fig5} the value $\tau_{D_B}\approx 0.17$ ps for $\theta_{\vec{k_c}}=0$. The expected value would be $\tau_{D_B} \approx 0.15$ ps with $v_f=10^6$ m/s and $d=154$ nm. The difference occurs since there are electrons described by the wave packet with a velocity slower than the Fermi velocity. 

The Bohmian dwell time $\tau_{D_B}$ increases with the increase in the angle of incidence until the critical angle $\theta_C$. This occurs because when increasing the incident angle, the angle at which the trajectory enters into the barrier also increases following the condition \eref{snell}. The effective distance that the electron has to traverse under the barrier is $d_{eff}=d/ \cos(\theta_{\vec{k_b}})$. See \fref{fig6}(b) for a definition of such a distance. The Bohmian dwell time can be written as: 
\begin{equation}
\tau_{D_B}=\frac{d \sqrt{E^2+V_0^2-2V_0E}} {v_f \sqrt{E^2\cos^2(\theta_{\vec{k_c}})+V_0^2-2V_0E}}
\label{t2}
\end{equation}
We notice that \eref{t2} reproduces \eref{t1} for $\theta_{\vec{k_c}}=0$. For the critical angle $\theta_{\vec{k_c}}=\theta_C$, the value of \eref{t2} gives infinite which means that the electron travels in the perpendicular direction $z$ inside the barrier, never reaching $x=b$. On the other hand, when $\theta_{\vec{k_c}} > \theta_C$ the number of transmitted electrons decreases, so most of the trajectories are either reflected from the barrier boundary, $x=a$, or are reflected after spending some time in the barrier and the estimation of the tunneling time then is more complex. In any case, there are very few electrons with $\theta_{\vec{k_c}} > \theta_C$.

\begin{figure}
\centering
\includegraphics[scale=0.3]{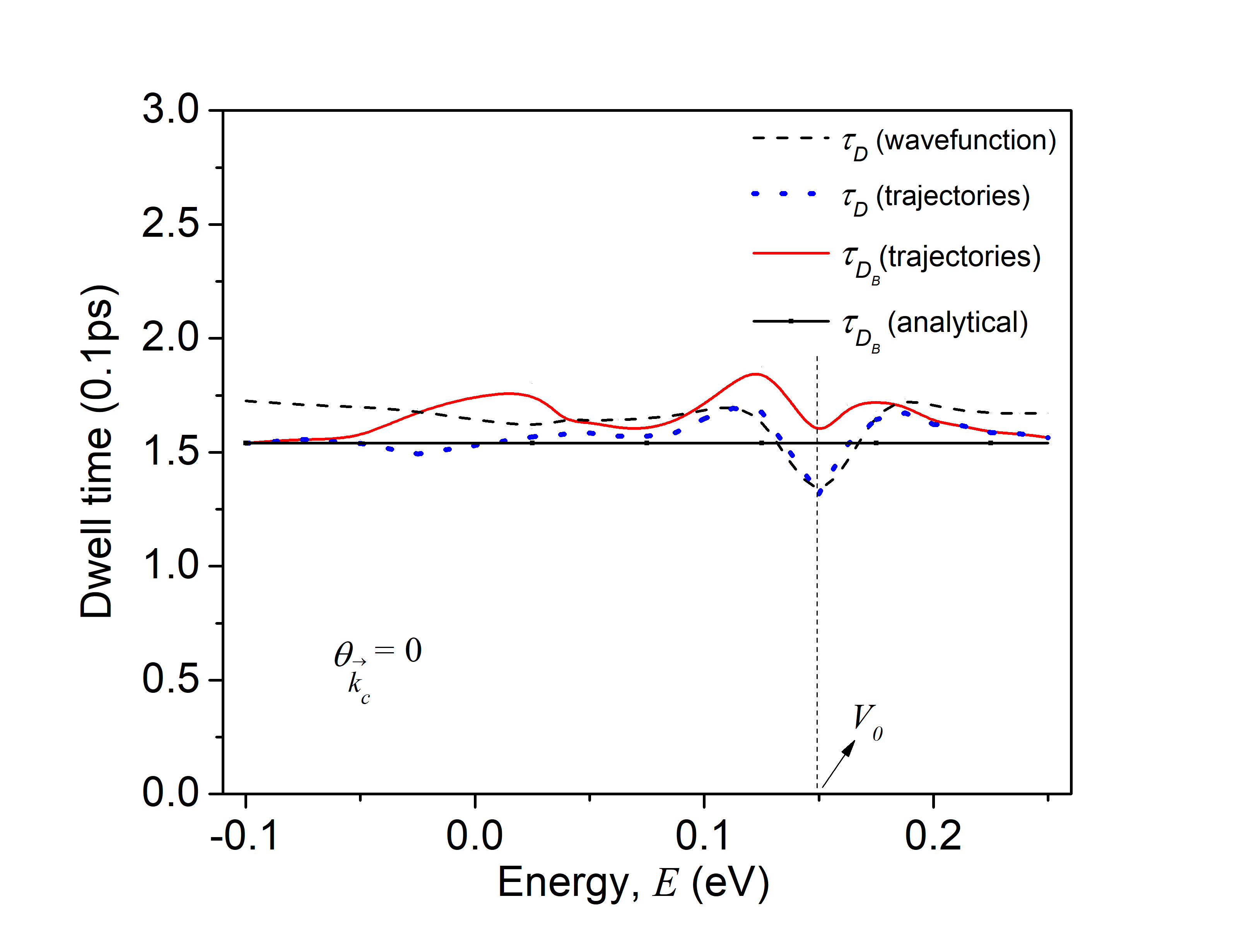}
\caption{Dwell time as a function of the positive and negative energy of electrons that impinge perpendicularly to the barrier computed from Eq.\eref{tun_orth} (black dashed line), Eq.\eref{tun_boh3} (blue dotted line), Eq.\eref{tun_boh7} (red solid line) and Eq.\eref{t2} (black solid line with square symbols).}
\label{fig4}
\end{figure}

Once we have analyzed the dependence of the orthodox and Bohmian dwell times on the incident angle, $\theta_{\vec{k_c}}$, let us discuss its dependence on the positive or negative kinetic energy $E$ for a zero incident angle $\theta_{\vec{k_c}}=0$. The main feature present in \fref{fig4} is that all electrons have similar Bohmian dwell time, roughly given by \eref{t1}, meaning that all electrons are moving with the Fermi velocity, $v_f=10^6$ m/s. 

\begin{figure}[!htbp]
\centering
\includegraphics[scale=0.3]{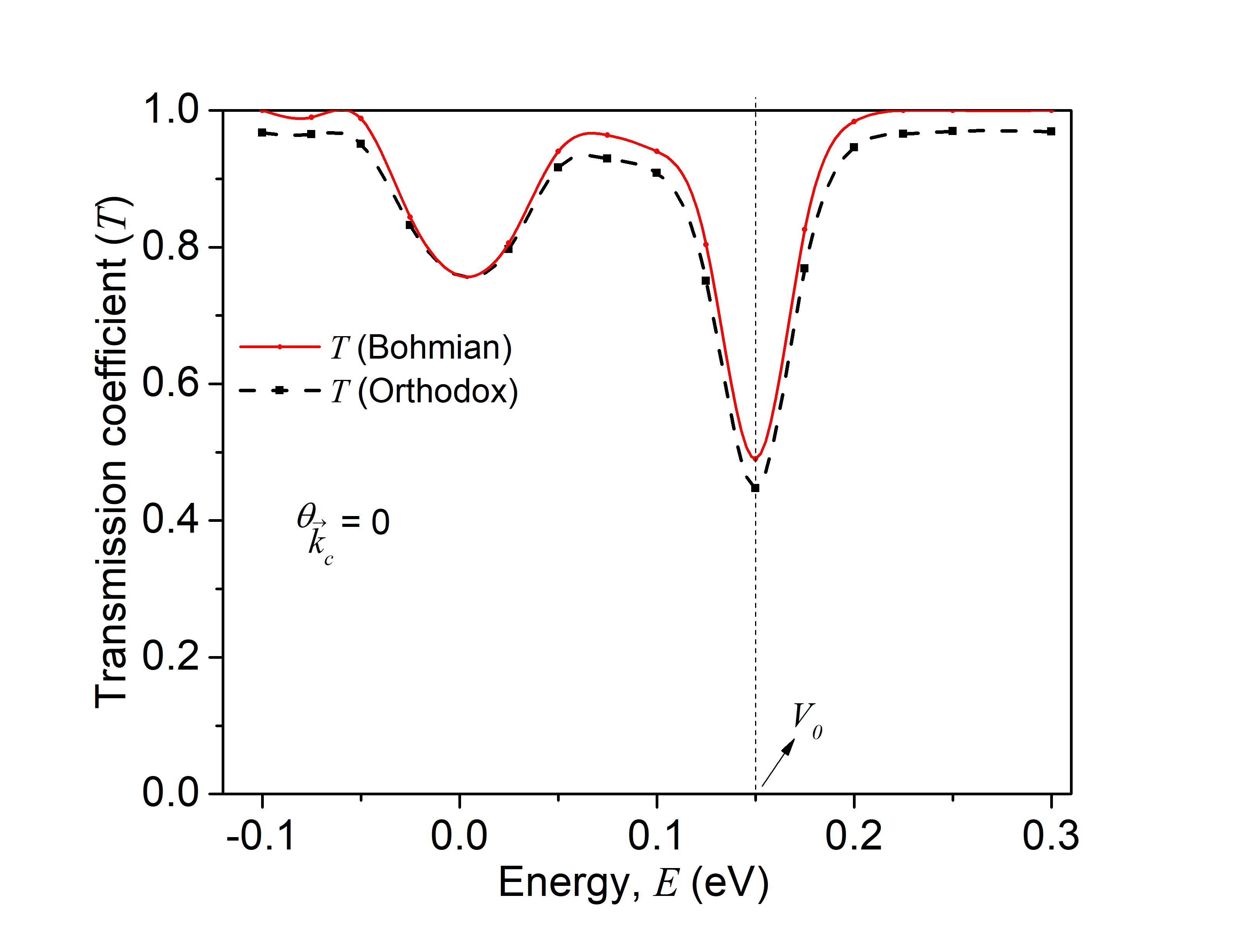}
\caption{Transmission coefficient for electrons that impinge perpendicularly to the barrier as a function of their initial energy.}
\label{fig4bis}
\end{figure}

\begin{figure}
\centering
\includegraphics[scale=0.3]{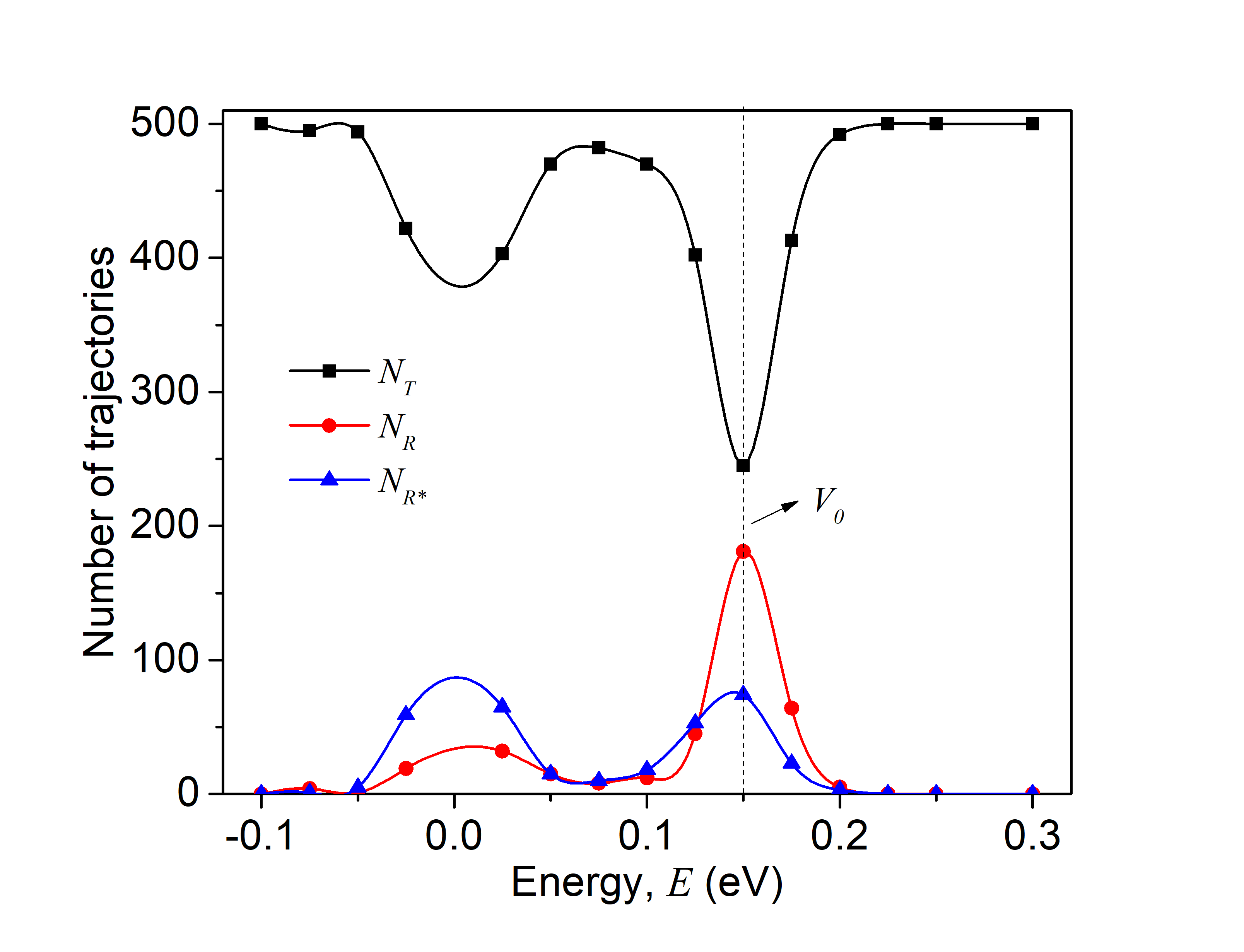}
\caption{Number of trajectories belonging to each of the three cases ($N_T$, $N_R$ and $N_{R^*}$) for electrons that impinge perpendicularly to the barrier as a function of their initial energy.}
\label{fig4bisbis}
\end{figure}

These results of the Klein tunneling in \fref{fig4} are in a great contradiction with what is usually found in semiconductor structures with parabolic band energies, where the dwell times strongly depends on the difference between the barrier and the electron energies, $V_0-E$. Here, even for negative kinetic energies, for example $E=-0.1$ eV,  or positive energies above the barrier, for example, $E=0.2$ eV, the predicted value of the dwell time given by the Fermi velocity is not greatly modified. In \fref{fig4bis} we have plotted the transmission coefficient given by \eref{T} (dashed line) and by \eref{Tb} (solid line), with great agreement. In \fref{fig4bisbis} we plot the number of particles $N_T$, $N_R$ and $N_{R^*}$ discussed in \sref{sec3}. The low number of reflected particles without even reaching the barrier region, $N_{R^*} \approx 0$, explains why the dwell times in \fref{fig4} are all almost identical. Only, small deviations are seen around $E \approx V_0 = 0.15$ eV and around $E \approx 0$. The first deviations around $E \approx V_0 = 0.15$ eV are explained by the effects of the conservation of the $z$ momentum shown in \fref{fig6}(a). The later deviations in both figures around $E \approx 0$ are mainly related to the difficulties of defining an initial wave packet around the Dirac Point. Because of the momentum uncertainty, such initial wave packet have positive and negative energies simultaneously.      

\section{Conclusion}
\label{sec5}

Motivated by the expected ability of graphene transistors to work at THz frequencies and the development of prototypes of graphene field effect transistors for high-frequency applications based on Klein tunneling phenomena \cite{dev1,dev2,dev3,dev4}, an analysis on the Klein tunneling times in graphene structures has been presented in this work. In particular, we study dwell times for electrons in a two-terminal graphene barrier using the BITLLES simulator \cite{bits}. We show that Bohmian trajectories are well suited formalism to discuss transit (tunneling) times and its relation to the cut-off frequencies of electron devices. 

 We have shown that Klein tunneling time (in gapless graphene with linear band structure) is not like the typical tunneling time (in materials with parabolic bands and with an energy gap). Such differences directly imply completely opposite features in the transit (tunneling) times of graphene structures in comparison to what is expected from traditional semiconductor structures with parabolic bands. 
 
 The main conclusions plotted in the figures of the text are next summarized. Because of the well known Klein paradox \cite{klein,nature}, for an incident angle equal to zero, $\theta_{\vec{k_c}}=0$, the transmission coefficient is roughly equal to the unity, $T=1$, with $N_R \approx 0$ and  $N_{R^*} \approx 0$. Then, the velocity of particles in the barrier region and outside is roughly equal to the Fermi velocity, $v_f=10^6$ m/s. This is true for all incident kinetic energy (with positive or negative kinetic energy). Then, the dwell time in the barrier region can be identically computed from the orthodox expression $\tau_D$ or the Bohmian one $\tau_{D_B}$, roughly estimated as $\tau_{D} \approx \tau_{D_B}  \approx d /v_f$.

For incident angles different from zero and smaller than the critical angle, $0<\theta_{\vec{k_c}} < \theta_C$, the transmission coefficient decreases because $N_{R^*} > 0$, but $N_R \approx 0$. Under these scenarios, the dwell time of the electrons has to be estimated only for the trajectories that spend some time in the barrier (what we name $N_T$ and $N_R$ in the text) but not by the trajectories $N_{R^*}$ that do not spend time in the barrier. Then, the orthodox expression $\tau_D$ in \eref{tun_orth} is not adequate and it has to be substituted by the Bohmian dwell time expression $\tau_{D_B}$. The dwell time can be roughly estimated as $\tau_{D_B} \approx d/ cos(\theta_{\vec{k_c}})/v_f$ where $d/cos(\theta_{\vec{k_c}})$ is the distance traversed by an electron in the barrier because of the Snell's law-like equation in \eref{snell}. Notice that $\tau_{D_B}$ is not a transmitted time, because it is not related with the transmitted particles $N_T$ only, but with $N_T$ and $N_R$, excluding $N_{R^*}$.

Finally, for incident angles larger than the critical angle, $0<\theta_{\vec{k_c}} > \theta_C$, the Bohmian dwell time can be computed numerically from $\tau_{D_B}$, but there is no simple expression for its evaluation because $N_R > N_{T}$ and it is not obvious what is the dwell time for the $N_R$ particles. Again, the Bohmian dwell time is different from the orthodox $\tau_D$ because the latter includes particles that do not enter into the barrier region ($N_{R^*} > 0$). In any case, there are few electrons traversing the barrier with such angles.

The main conclusion of this work, regarding the high-frequency capabilities of graphene devices based on Klein tunneling is that the high graphene mobility is roughly independent of the presence of Klein tunneling phenomena in the active device region.  The reason is a direct consequence of the graphene band structure. All electrons, at all positive or negative kinetic energies, move roughly with the Fermi velocity above or below the potential barrier.

At this point we want to notice that the relation between cut-off frequencies and tunneling time has been analyzed in \sref{sec:2bis} for an idealized two terminal device under the assumption that $L_x< L_z, L_y$. In more general scenarios, for example in a three terminal device, like a graphene transistor, the expression \eref{2terminal} for the current is no longer valid. This means that the cut-off frequencies cannot be directly linked to the inverse of the transit (tunneling) time. Further discussion of this issue can be found in \cite{Zhen_TED}.

Apart from the previous conclusions devoted to graphene devices for high-frequency applications, we have a final remark on the type of simulators required for predicting high-frequency features of nanoelectronic devices. There are several quantum theories empirically equivalent to discuss the quantum behavior of nanoelectronic devices at high-frequency. By construction, the Bohmian theory \cite{unmeasured} has the ability to provide measured and unmeasured properties (for example, particle positions or the total current) for quantum systems. Such ability is very convenient because it allows to get information of the system that are not contaminated by the measurement. This is specially relevant in the two consecutive measurements required to get transit (tunneling) times. Later, if we know how to relate measured and unmeasured properties in one experimental set-up (such as the high-frequency measurement set-up defined in \cite{PRL damiano}), the unmeasured properties provided by the Bohmian theory become very useful for computations of high-frequency applications of electronic devices. The BITLLES simulator used in this work is a clear example of this computing strategy.  

\section*{Acknowledgment}
\addcontentsline{toc}{section}{Acknowledgment}
We are grateful to Dmitry Sokolovski for enlightening discussions on tunneling times. We acknowledge funding from Fondo Europeo de Desarrollo Regional (FEDER), the ``Ministerio de Ciencia e Innovaci\'{o}n'' through the Spanish Project TEC2015-67462-C2-1-R, the Generalitat de Catalunya (2014 SGR-384), the European Union's Horizon 2020 research and innovation programme under grant agreement No Graphene Core2 785219 and under the Marie Skłodowska-Curie grant agreement No 765426 (TearApps). 

\appendix

\section{Graphene electron trajectories}
\label{GrTr}
Here we demonstrate how the Bohmian velocity \eref{bvel} can be obtained from the Dirac equation \eref{Dirac}. The typical procedure to deduce the Bohmian velocity from the Schr\"{o}dinger equation gives:
\begin{equation}
\vec{v}(\vec{r},t)=\frac{d\vec{r}}{dt}=\frac{\hbar}{m}\operatorname{Im}\left(\frac{\nabla\psi(\vec{r})}{\psi(\vec{r})}\right)
\end{equation}
We adapt here the previous procedure to deduce the Bohmian velocity associated to the Dirac equation (for details see \cite{kike}). To obtain the current density, we find out first the continuity equation from the Dirac equation \eref{Dirac} rewritten here as:
\begin{equation}
i\hbar\dfrac{\partial\psi(\vec{r},t)}{\partial t}=-i\hbar v_f(\vec{\sigma}\cdot\vec{\triangledown})\psi(\vec{r},t)
\label{Dirac0}
\end{equation}
with $\vec r=\{x,z\}$. Now, multiply the Hamiltonian by the conjugated wave function:
\begin{equation}
\psi(\vec{r},t)^{\dagger}i\hbar\dfrac{\partial\psi(\vec{r},t)}{\partial t}=-i\psi(\vec{r},t)^{\dagger}\hbar v_f(\vec{\sigma}\cdot\vec{\triangledown})\psi(\vec{r},t)
\label{Dirac1}
\end{equation}
and conjugate and transpose the above equation \eref{Dirac1}:
\begin{equation}
\psi(\vec{r},t)\dfrac{\partial\psi(\vec{r},t)^{\dagger}}{\partial t}=-\psi(\vec{r},t) v_f(\vec{\sigma}\cdot\vec{\triangledown})\psi(\vec{r},t)^{\dagger}
\label{Dirac2}
\end{equation}
The above expression \eref{Dirac2} implies that the Pauli matrices are hermitian. If we now add \eref{Dirac1} and \eref{Dirac2} we get:
\begin{eqnarray}
\psi(\vec{r},t)^{\dagger}\dfrac{\partial\psi(\vec{r},t)}{\partial t}+\psi(\vec{r},t)\dfrac{\partial\psi(\vec{r},t)^{\dagger}}{\partial t}\\{\nonumber}=-\left[\psi(\vec{r},t)^{\dagger} v_f(\vec{\sigma}\cdot\vec{\triangledown})\psi(\vec{r},t)+\psi(\vec{r},t) v_f(\vec{\sigma}\cdot\vec{\triangledown})\psi(\vec{r},t)^{\dagger}\right]
\label{Dirac3}
\end{eqnarray}
which leads directly to the continuity equation:
\begin{equation}
\dfrac{\partial|\psi(\vec{r},t)|^2}{\partial t}+\vec{\triangledown}\cdot \left( v_f\psi(\vec{r},t)^{\dagger}\vec{\sigma}\psi(\vec{r},t)\right)=0
\label{ContinuityEquation}
\end{equation}
where we can easily identify the probability current density (of the Dirac equation) as:
\begin{equation}
\vec J(\vec{r},t)= v_f\psi(\vec{r},t)^{\dagger}\vec{\sigma}\psi(\vec{r},t)
\label{current}
\end{equation}
From here, we can also identify the Bohmian velocities by using the general expression $\vec J(\vec{r,t)}=\rho \vec{v}=|\psi(\vec{r},t)|^2 \vec{v}$ in \eref{ContinuityEquation} so that ,
\begin{equation}
\vec{v}(\vec{r},t)=\dfrac{J(\vec{r},t)}{|\psi(\vec{r},t)|^2}=\dfrac{ v_f\psi(\vec{r},t)^{\dagger}\vec{\sigma}\psi(\vec{r},t)}{|\psi(\vec{r},t)|^2}
\label{bvel1}
\end{equation}
And from the above equation the bohmian velocity in the $x$ and $z$ directions can be given as :
\begin{equation}
v_x(\vec{r},t)=\dfrac{J_x(\vec{r},t)}{|\psi(\vec{r},t)|^2}=\dfrac{ v_f\psi(\vec{r},t)^{\dagger}\sigma_x\psi(\vec{r},t)}{|\psi(\vec{r},t)|^2}
\label{vbx}
\end{equation}
and,
\begin{equation}
v_z(\vec{r},t)=\dfrac{J_z(\vec{r},t)}{|\psi(\vec{r},t)|^2}=\dfrac{ v_f\psi(\vec{r},t)^{\dagger}\sigma_z\psi(\vec{r},t)}{|\psi(\vec{r},t)|^2}
\label{vby}
\end{equation}
Since \eref{vbx} and \eref{vby} are independent of $s$ (see \eref{bispinorini} in the text), it is evident that independently of whether the electrons are in the conduction or valence band, they move in the same direction. It is important to emphasize that the identity $\vec J(\vec{r,t)}=\rho \vec{v}=|\psi(\vec{r},t)|^2 \vec{v}$ in \eref{ContinuityEquation} used above guarantees the empirical equivalence between orthodox and Bohmian mechanics.

\end{document}